\documentclass[useAMS,usenatbib]{mn2e}

\usepackage{latexsym,graphicx,natbib}

\usepackage{color}

\newcommand\kms{{\rm\,km\,s^{-1}}}

\def\apgt{\ {\raise-.5ex\hbox{$\buildrel>\over\sim$}}\ }
\def\aplt{\ {\raise-.5ex\hbox{$\buildrel<\over\sim$}}\ }

\title[MN112: a new Galactic candidate Luminous Blue Variable]{MN112: a new Galactic candidate Luminous Blue Variable\footnotemark[0]\thanks{
Partially based on observations collected at the German-Spanish Astronomical Center,
Calar Alto, jointly operated by the Max-Planck-Institut f\"ur Astronomie
Heidelberg and the  Instituto de Astrof\'isica de Andaluc\'ia (CSIC).}}
\author[V.V.Gvaramadze et al.]
       {V. V.~Gvaramadze,$^{1}$\thanks{E-mail: vgvaram@mx.iki.rssi.ru (VVG); akniazev@saao.ac.za (AYK);
       fabrika, olga@sao.ru (SF, OS); berdnik, cher, alla@sai.msu.ru (LNB, AMC, AVZ)},
       A. Y.~Kniazev,$^{2,3}$, S.~Fabrika,$^{4}$ O.~Sholukhova,$^{4}$
        \newauthor
        L. N.~Berdnikov,$^{1,5}$ A. M.~Cherepashchuk,$^{1}$ and A. V.~Zharova$^{1}$\\
        $^{1}$Sternberg Astronomical Institute, Moscow State University, Universitetskij Pr. 13, Moscow 119992, Russia\\
        $^{2}$South African Astronomical Observatory, PO Box 9, 7935 Observatory, Cape Town,
        South Africa \\
        $^{3}$Southern African Large Telescope Foundation, PO Box 9, 7935 Observatory, Cape Town,
        South Africa \\
        $^{4}$Special Astrophysical Observatory, Nizhnij Arkhyz, 369167, Russia\\
        $^{5}$Isaac Newton Institute of Chile, Moscow Branch, Universitetskij Pr. 13, Moscow 119992, Russia
        }
\begin{document}

\date{Accepted 2010 February 2. Received 2010 February 2; in original form 2008 December 18}


\maketitle

\label{firstpage}

\begin{abstract}

We report the discovery of a new Galactic candidate Luminous Blue
Variable (cLBV) via detection of an infrared circular nebula and
follow-up spectroscopy of its central star. The nebula, MN112, is
one of many dozens of circular nebulae detected at $24 \, \mu$m in
the {\it Spitzer Space Telescope} archival data, whose morphology is
similar to that of nebulae associated with known (c)LBVs and related
evolved massive stars. Specifically, the core-halo morphology of
MN112 bears a striking resemblance to the circumstellar nebula
associated with the Galactic cLBV GAL\,079.29+00.46, which suggests
that both nebulae might have a similar origin and that the central
star of MN112 is a LBV. The spectroscopy of the central star showed
that its spectrum is almost identical to that of the bona fide LBV
P\,Cygni, which also supports the LBV classification of the object.
To further constrain the nature of MN112, we searched for signatures
of possible high-amplitude ($\ga 1$ mag) photometric variability of
the central star using archival and newly obtained photometric data
covering a 45 year period. We found that the B magnitude of the star
was constant ($\simeq$ 17.1$\pm$0.3 mag) over this period, while in
the I band the star brightened by $\simeq 0.4$ mag during the last
17 years. Although the non-detection of large photometric
variability leads us to use the prefix `candidate' in the
classification of MN112, we remind that the long-term photometric
stability is not unusual for genuine LBVs and that the brightness of
P\,Cygni remains relatively stable during the last three centuries.
\end{abstract}

\begin{keywords}
line: identification -- circumstellar matter -- stars:
emission-line, Be
\end{keywords}

\section{Introduction}
\label{sec:intro}

Luminous Blue Variables (LBVs) are rare evolved massive stars in the
upper left of the HR diagram (Conti 1984). Besides the very high
luminosity, most LBVs are characterized by irregular large
spectrophotometric variability (Humphreys \& Davidson 1994; van
Genderen 2001), which manifests itself in drastic changes in
spectral type and visual brightness (by 1-2 mag) of these objects.
The origin of the variability and its characteristic timescale are
not yet well understood. Although the majority of confirmed LBVs
show variations over decades or even years, some of them (e.g.
P\,Cygni) are photometrically stable during the much longer periods
of time (e.g. de Groot, Sterken \& van Genderen 2001).

It is believed that LBVs represent an intermediate phase in the
evolution of the most massive stars from the core-hydrogen burning O
stars to the hydrogen-poor Wolf-Rayet stars, during which a massive
star loses a significant fraction of its initial mass through the
copious stellar wind or in the form of instant outbursts. As a
result of this mass loss, most, if not all, of established and
candidate LBVs (cLBVs) are surrounded by compact ($\la 1$ pc)
circumstellar nebulae (e.g. Weis 2001; Clark, Larionov \& Arkharov
2005), detected either though direct imaging (in the optical,
infrared or radio wavebands) or inferred via the presence of
forbidden emission lines in their spectra. These nebulae show a wide
range of morphologies, ranging from simple round to bipolar and
triple-ring forms (e.g. Nota et al. 1995; Weis 2001; Smith 2007;
Gvaramadze, Kniazev \& Fabrika 2009a).

Detection of LBV-like nebulae around stars with P\,Cygni-type or
$\eta$\,Car-type spectra could serve as circumstantial evidence that
these stars are genuine LBVs even if their photometric variability
has not yet been detected (Bohannan 1997; Massey et al. 2007). On
the other hand, search for LBV-like nebulae accompanied by
spectroscopic follow-up of their central stars could be a powerful
tool for detection of new (c)LBVs and related evolved stars (Clark
et al. 2003; Gvaramadze et al. 2009a,b,c).

In this Letter, we report the discovery of a new Galactic cLBV via
detection of a ring-like nebula (reminiscent of the circumstellar
nebula of the cLBV GAL 079.29+00.46) and follow-up spectroscopy of
its central star, revealing a striking similarity between the
spectrum of the star and that of the bona fide LBV P\,Cygni.

\section{Infrared nebula MN112 and its central star}
\label{sec:nebula}

The new Galactic cLBV was identified via detection of its
circumstellar nebula (Gvaramadze et al. 2009a) in the {\it Spitzer
Space Telescope} archival data obtained with the Multiband Imaging
Photometer for {\it Spitzer} (MIPS; Rieke et al. 2004) within the
framework of the 24 and 70 Micron Survey of the Inner Galactic Disk
with MIPS (MIPSGAL; Carey et al. 2009). This nebula is one of many
dozens of circular nebulae discovered at MIPS $24 \, \mu$m images,
whose appearance is very similar to that of nebulae associated with
known (c)LBVs and related evolved massive stars (Gvaramadze et al.
2009a).
\begin{figure}
\begin{center}
\includegraphics[width=1.0\columnwidth,angle=0]{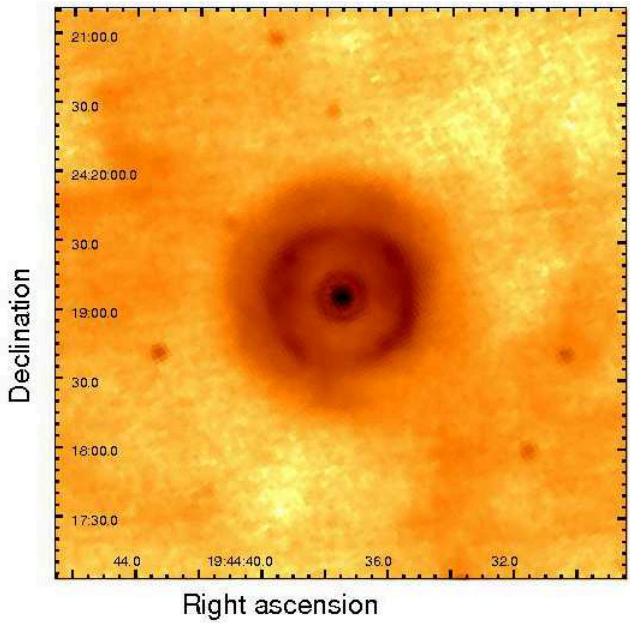}
\end{center}
\caption{
{\it Spitzer} MIPS $24 \, \mu$m image of the ring nebula MN112 and its central star.
    }
\label{fig:neb}
\end{figure}
Fig.\,\ref{fig:neb} shows the MIPS $24 \, \mu$m image of the nebula
[hereafter MN112, in accordance with the nomenclature adopted in
Gvaramadze et al. (2009a)] and its central star. MN112 consists of a
bright circular shell (with a radius of $\simeq 33$ arcsec)
surrounded by a halo (with a radius of $\simeq 54$ arcsec). The
core-halo morphology of MN112 strikingly resembles that of the
circumstellar nebula of the Galactic cLBV GAL 079.29+00.46
(Gvaramadze et al. 2009a). This similarity suggests that both
nebulae might have a similar nature and that the central star of
MN112 is a (c)LBV as well.

MN112, like the majority of other LBV-like nebulae detected with
{\it Spitzer}, is visible only at $24\,\mu$m (cf. Carey et al.
2009). Unlike the nebula, the central star (in what follows, we will
use for the star the same name as for the nebula) is visible not
only at $24\,\mu$m but also at all four (3.6, 4.5, 5.8 and
$8.0\,\mu$m) images obtained with the {\it Spitzer} Infrared Array
Camera (IRAC; Fazio et al. 2004) within the Galactic Legacy Infrared
Mid-Plane Survey Extraordinaire (GLIMPSE; Benjamin et al. 2003).
According to the GLIMPSE\,I Spring\,'07 Archive\footnote{Available
at http://irsa.ipac.caltech.edu/index.html}, the star is located at
$\alpha_{2000} =19^{\rm h} 44^{\rm m} 37\fs60, \delta_{J2000}
=24\degr 19\arcmin 05\farcs9$ and has the following IRAC magnitudes:
[3.6]=6.73$\pm$0.03, [4.5]=6.42$\pm$0.04, [5.8]=6.18$\pm$0.02,
[8.0]=5.84$\pm$0.02. Using the IRAC photometry, 2MASS magnitudes
($J,H,K_{s}$)=(8.86$\pm$0.02, 8.02$\pm$0.02, 7.42$\pm$0.02) of the
star (Skrutskie et al. 2006) and the colour-colour diagrams by
Hadfield et al. (2007), one finds that MN112 falls in the region
populated by Wolf-Rayet and other hot, evolved massive stars.

\begin{table}
  \caption{Photometry of MN112.}
  \label{tab:phot}
  \renewcommand{\footnoterule}{}
  \begin{center}
  \begin{minipage}{\textwidth}
    \begin{tabular}{cccc}
      \hline
      Date & $B$ & $V$ & $I$ \\
      \hline
      1990 July 25$^{(1)}$ & 17.06$\pm$0.21 & -- & -- \\
      1992 July 23$^{(1)}$ & -- & -- & 11.56$\pm$0.12 \\
      1996 July 12$^{(2)}$ & 16.90$\pm$0.18 & -- & -- \\
      2009 April 22$^{(3)}$ & 17.13$\pm$0.12 & 14.53$\pm$0.03 & 11.15$\pm$0.03 \\
      2009 June 21$^{(4)}$ & 16.94$\pm$0.05 & 14.46$\pm$0.04 & -- \\
      \hline
    \end{tabular}
    \end{minipage}
    \end{center}
     (1) POSS-II; (2) GSC 2.2; (3) 40-cm Meade telescope; (4) 6-m/SCORPIO
\end{table}

The optical counterpart to MN112 was identified by Dolidze (1975) as
an emission-line star, named in the SIMBAD database as
[D75b]\,Em*\,19-008\footnote{Note that the SIMBAD database provides
erroneous coordinates for the star.}. This star was included in the
Catalogue of H-alpha emission stars in the northern Milky Way
(Kohoutek \& Wehmeyer 1999). Two other SIMBAD names of the star are
[KW97]\,44-47 and HBHA\,4202-22.

We determined the photometric B, V and I magnitudes of MN112 on CCD
frames obtained with the SBIG CCD ST-10XME attached to the 40-cm
Meade telescope of the Cerro Armazones Astronomical Observatory of
the Northern Catholic University (Antofagasta, Chile) during our
observations in 2009 April 22. The results are given in
Table\,\ref{tab:phot}, where we also present photometry obtained
from different sources and calibrated using the secondary
photometric standards established from the Meade data (see
Section\,\ref{sec:LBV} for more details).

\begin{figure*}
\begin{center}
\includegraphics[width=12cm,angle=270,clip=]{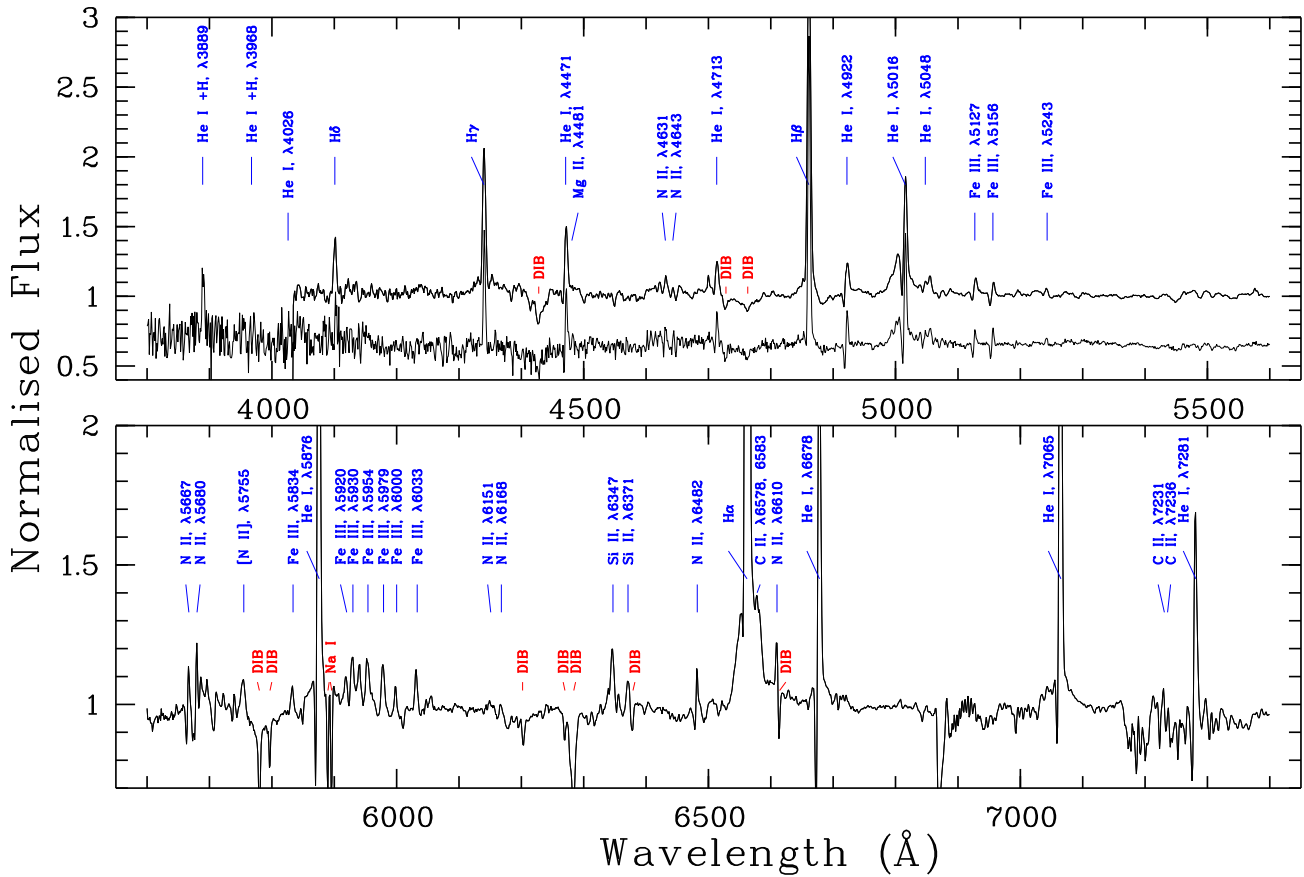}
\end{center}
\caption{Normalized spectra of MN112 with principal lines and most
prominent DIBs indicated. Top panel: blue spectra taken with the
TWIN spectrograph and the SCORPIO focal reducer (the TWIN spectrum
is shifted downwards by 0.3 continuum flux unit). Bottom panel: red
spectrum taken with the TWIN spectrograph. } \label{fig:spec}
\end{figure*}

\section{Spectroscopic follow-up}
\label{sec:observ}

To determine the nature of MN112, we obtained spectra of the central
star with the Cassegrain Twin Spectrograph (TWIN) of the 3.5-m
telescope in the Observatory of Calar Alto (Spain) during director's
discretionary time on 2009 May 5 (exposure time of $3\times600$
sec). The setup used for TWIN was the grating T08 in the first order
for the blue arm (spectral range 3500--5600 \AA) and T04 in the
first order for the red arm (spectral range 5300--7600 \AA) which
provides an inverse dispersion of 72~\AA/mm for both arms. Slit of
$240\arcsec \times2.1\arcsec$ was used for these spectral
observations. The resulting FWHM spectral resolution measured on
strong lines of night sky and reference spectra was 3--3.5~\AA. The
seeing during the observations was stable, $\simeq 1.3-1.4$ arcsec.

Additional spectra were obtained with the Russian 6-m telescope
using the SCORPIO\footnote{Spectral Camera with Optical Reducer for
Photometrical and Interferometrical Observations;
http://www.sao.ru/hq/lsfvo/devices/scorpio/scorpio.html} focal
reducer (Afanasiev \& Moiseev 2005) in a long-slit mode with a slit
width of 1 arcsec, providing a spectral resolution of 5 \AA. The
spectra were taken on 2009 June 21 in spectral ranges
$\lambda\lambda 4030-5830$~\AA \, (exposure time of 2400 sec) and
$\lambda\lambda 5730-7500$~\AA \, (exposure time of 1800 sec). The
seeing during observations was $\simeq 1.5$ arcsec.

Data reduction was performed using the standard procedures. The
resulting spectra are shown in Fig.\,\ref{fig:spec}. The top panel
presents blue spectra taken with both telescopes, as the TWIN
spectrum has better resolution, while the SCORPIO one has better
signal-to-noise ratio. The bottom panel shows the TWIN spectrum in
the red region, where the TWIN spectrograph provides better
resolution. Detailed comparison of the spectra from both telescopes
does not show any evidence of significant change.

Equivalent widths (EWs), FWHMs and radial heliocentric velocities
(RVs) of main emission lines [measured applying the MIDAS programs;
see Kniazev et al. (2004) for details] are summarized in
Table\,\ref{tab:int}. For EWs we give their mean values derived from
both spectra, while for measurements of FWHMs and RVs we used the
TWIN spectrum alone (owing to its better resolution).

\section{MN112 -- a new candidate LBV}
\label{sec:LBV}

\begin{figure}
\begin{center}
\includegraphics[width=6cm,angle=270]{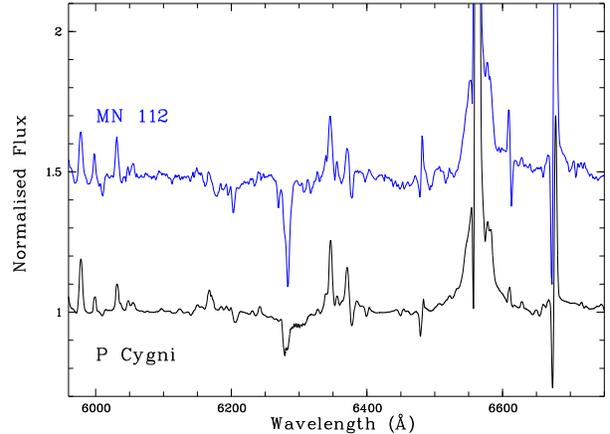}
\end{center}
\caption{Comparison of a portion of the TWIN spectrum of MN112 (shifted upwards
by 0.5 continuum flux unit)
  with the corresponding part of the P\,Cygni spectrum (Stahl et al. 1993).
  The spectrum of P\,Cygni is degraded to the resolution of the TWIN spectrum.
  The higher extinction towards MN112 is manifested in much
  stronger DIBs (e.g. at 6280 and 6613 \AA).}
  \label{fig:twospec}
\end{figure}

The first look on Fig.\,\ref{fig:spec} shows that the spectrum of
MN112 is extremely similar to that of the bona fide LBV P\,Cygni
(see Fig.\,\ref{fig:twospec}). We therefore used the spectral atlas
of P\,Cygni by Stahl et al. (1993) to identify the lines indicated
in Fig.\,\ref{fig:spec}. The spectrum of MN112, like P\,Cygni, is
dominated by strong emission lines of hydrogen and He\,{\sc i}. The
Balmer lines have prominent wings and weak P\,Cygni absorption
components, while the He\,{\sc i} lines show strong P\,Cygni
profiles.

Further emission lines in the spectrum of MN112 are numerous
permitted lines of N\,{\sc ii}, Fe\,{\sc iii} and Si\,{\sc ii}. The
dichotomy of Fe\,{\sc iii} lines observed in the spectrum of
P\,Cygni (Stahl et al. 1993) is present in the spectrum of MN112 as
well: lines with low multiplet numbers (e.g. Fe\,{\sc iii}
$\lambda\lambda 5127, 5156$) have distinct P Cygni profiles, while
those with higher multiplet numbers (e.g. Fe\,{\sc iii}
$\lambda\lambda 5243, 5920-6033$) are purely in emission. Some of
N\,{\sc ii} lines (e.g. $\lambda\lambda 6151, 6168$) are also purely
in emission. No He\,{\sc ii} emission lines are present in the
spectrum.

The only forbidden line detected in the spectrum of MN112 is the
line of [N\,{\sc ii}] $\lambda 5755$, although the presence of two
weak [N\,{\sc ii}] $\lambda\lambda 6548, 6583$ lines in the
H$\alpha$ emission wings cannot be excluded (cf. Markova \& de Groot
1997). The relative prominence of the line [N\,{\sc ii}] $\lambda
5755$ implies that [N\,{\sc ii}] emission originates in the dense
($\sim 10^7 \, {\rm cm}^{-3}$) matter, probably close to the star
(Stahl et al. 1991). The FWHM of the [N\,{\sc ii}] $\lambda 5755$
line could be used as a measure of the stellar wind velocity,
$v_\infty$ (e.g. Crowther, Hillier \& Smith 1995). Using FWHM from
Table\,\ref{tab:int}, one finds $v_\infty \simeq 400\pm40 \, \kms$,
which is about two times higher than that of P\,Cygni (e.g. Barlow
et al. 1994; Najarro, Hillier \& Stahl 1997).

\begin{table}
\caption{EW, FWHM and RV of the main emission lines in the spectrum
of MN112.} \label{tab:int}
\begin{center}
\begin{minipage}{\textwidth}
\begin{tabular}{ccccc} \hline
$\lambda _{0}$ & Ion & EW($\lambda$) & FWHM($\lambda$) & RV  \\
  (\AA) &   & (\AA) & (\AA) & ($\kms$) \\
\hline
3889 & He\ {\sc i}\ +\ H8\   & 3.8$\pm$0.4  & 4.4$\pm$0.4    & -- \\
4101 & H$\delta$\            & 1.5$\pm$0.1  & 1.6$\pm$0.5    & -- \\
4340 & H$\gamma$\            & 4.8$\pm$0.1  & 2.2$\pm$0.2    &--35$\pm$8  \\
4471 & He\ {\sc i}\          & 2.3$\pm$0.3  & 1.3$\pm$0.5    &  24$\pm$15 \\
4713 & He\ {\sc i}\          & 1.6$\pm$0.2  & 2.8$\pm$0.5    &  22$\pm$13 \\
4861 & H$\beta$\             &16.2$\pm$0.3  & 3.4$\pm$0.2    &--23$\pm$7  \\
4922 & He\ {\sc i}\          & 1.2$\pm$0.2  &     --         &  40$\pm$20 \\
5016 & He\ {\sc i}\          & 4.6$\pm$0.3  & 2.7$\pm$0.5    &  9$\pm$12 \\
5127 & Fe\ {\sc iii}\        & 0.5$\pm$0.2  & 1.4$\pm$1.2    & -- \\
5156 & Fe\ {\sc iii}\        & 0.5$\pm$0.2  &     --         & -- \\
5755 & [N\ {\sc ii}]\        & 1.1$\pm$0.2  & 7.6$\pm$0.7    &--65$\pm$9  \\
5834 & Fe\ {\sc iii}\        & 0.6$\pm$0.2  & 5.3$\pm$0.4    &--65$\pm$8  \\
5876 & He\ {\sc i}\          &14.7$\pm$0.4  & 3.6$\pm$0.3    &--7$\pm$6  \\
5979 & Fe\ {\sc iii}\        & 1.1$\pm$0.2  & 5.1$\pm$0.3    &--56$\pm$6  \\
6000 & Fe\ {\sc iii}\        & 0.4$\pm$0.1  & 4.0$\pm$0.4    &--68$\pm$9  \\
6033 & Fe\ {\sc iii}\        & 0.8$\pm$0.2  & 4.2$\pm$0.3    &--66$\pm$7  \\
6347 & Si\ {\sc ii}\         & 1.6$\pm$0.2  & 8.6$\pm$0.7    &--75$\pm$15 \\
6371 & Si\ {\sc ii}\         & 0.7$\pm$0.2  & 4.3$\pm$0.7    &--28$\pm$14 \\
6482 & N\ {\sc ii}\          & 0.7$\pm$0.2  & 2.8$\pm$0.6    &  21$\pm$12 \\
6563 & H$\alpha$\            &64.1$\pm$0.8  & 5.8$\pm$0.2    &--16$\pm$5  \\
6678 & He\ {\sc i}\          & 7.5$\pm$0.3  & 3.2$\pm$0.3    &  7$\pm$8  \\
7065 & He\ {\sc i}\          &12.2$\pm$0.3  & 4.2$\pm$0.2    &--20$\pm$5  \\
7281 & He\ {\sc i}\          & 4.5$\pm$0.3  & 4.1$\pm$0.5    &  5$\pm$8  \\
\hline
\end{tabular}
\end{minipage}
\end{center}
\end{table}

Numerous diffuse interstellar bands (DIBs) are present in the
spectrum. In the blue region they are $\lambda\lambda 4428, 4726$
and 4762, in the red the strongest DIBs are at 5780, 5797, 6280,
6379 and 6613 \AA. Strong Na\,{\sc i} absorption lines detected in
the red spectrum are of circumstellar or interstellar origin, while
the strong absorptions visible at $\lambda > 6800$ \AA \, are
telluric.

Overall, the spectrum of MN112 is extremely similar to that of the
bona fide LBV P\,Cygnus (Stahl et al. 1993), except for the lack of
strong P\,Cygni absorptions in the Balmer lines. In this connection
it is worthy of note that the P\,Cygni profiles of Balmer lines in
spectra of some (c)LBVs can disappear and reappear again over
time-scales of several years or decades (e.g. Crowther 1997; Massey
et al. 2007) or even weeks (Smith, Crowther \& Prinja 1994). A
possible explanation of this behaviour is that the P\,Cygni
absorptions arise in the recombined hydrogen in the outermost layers
of the stellar wind and disappear when the stellar wind is fully
ionized (Crowther 1997; Najarro et al. 1997). The changes in the
ionization state of the absorbing material might be caused by
variability of the wind density, which in turn is due to variability
of the stellar mass-loss rate and/or the wind velocity.

We now turn to estimates of the reddening towards MN112. Proceeding
from the close similarity between the spectra of MN112 and P Cygni,
it is tempting to assume that both stars have the same temperatures
and therefore the same intrinsic colours. For $B=17.13$ mag and
$V=14.53$ mag (see Table\,\ref{tab:phot}) and adopting from (Najarro
et al. 1997) the intrinsic colour of $(B-V)_0 =-0.26$ mag, one finds
$E(B-V)\simeq 2.86$ mag. An indirect support to this estimate comes
from comparison of dereddened spectral slopes of MN112 and P\,Cygni,
which should coincide with each other if the temperatures of both
stars are similar. For this comparison we obtained a spectrum of
P\,Cygni with SCORPIO on 2009 September 19. We found that the
SCORPIO spectra of both stars nicely match each other if one
dereddens the spectrum of MN112 by $\simeq 2.2$ mag in $E(B-V)$.
Adopting $E(B-V)=0.51$ mag for P\,Cygni (Najarro et al. 1997), one
finds $E(B-V)=2.71$ mag for MN112, which is in a good agreement with
the above estimate.

MN112 is located in the direction towards the OB association Vul
OB1 at $\simeq 1$ degree from the open cluster NGC\,6823 (which is
part of Vul\,OB1). The distance estimates to NGC\,6823 range from
$\simeq 2$ to 3.5 kpc (e.g. Sagar \& Joshi 1981; Kharchenko et al.
2005), which puts the cluster (and association) in the local
(Orion) spiral arm. Assuming that MN112 is a member of Vul OB1
association, one finds its absolute visual magnitude $M_V \simeq
-(5.8-7.1)$ mag and luminosity of $\log (L/L_{\odot} ) \simeq
4.9-5.4$ [here we assumed that the bolometric correction of MN112
is equal to that of P\,Cygni, $-1.54$ mag (Najarro, personal
communication)]. These estimates imply that, in principle, MN112
might belong to a group of low-luminosity (c)LBVs with $\log
(L/L_{\odot}) \simeq 5.2-5.6$ (Humphreys \& Davidson 1994; see
also Clark et al. 2009), provided that it is located at the
distance of $\sim 3.5$ kpc. Another possibility is that MN112 is
located in the next arm out, the Perseus Arm. An indirect support
of this possibility comes from our estimate of the visual
extinction towards MN112, $A_V=3.1E(B-V)\simeq 8.4-8.9$ mag, which
is several times larger than that towards Vul\,OB1 and its central
cluster NGC\,6823 [$2.3-3.3$ mag (The \& van Paradijs 1971) and
2.6 mag (Kharchenko et al. 2009), respectively]. In this case, the
distance to MN112 is $\ga 7$ kpc, which corresponds to $M_V \la
-8.6$ mag and $\log (L/L_{\odot} ) \ga 6.0$ [$M_V$ and $\log
(L/L_{\odot} )$ of P\,Cygni are equal to $-8.0$ mag and 5.8,
respectively; Najarro et al. 1997].

To further constrain the nature of MN112, we searched for its
possible large-scale ($\ga 1$ mag) photometric variability, using
the secondary photometric standards established with the 40-cm Meade
telescope. We used CCD B and V frames taken with the Russian 6-m
telescope and recalibrated photometry from POSS-II B and I plates
and the Guide Star Catalog 2.2 (McLean et al. 2000). These
measurements (summarized in Table\,\ref{tab:phot}) show that the B
magnitude of MN112 remains constant within error margins over the
last 19 years, while in the I band the star brightened by
0.41$\pm0.12$ mag. Also, we used the collection of photographic
plates of the Sternberg Astronomical Institute (Moscow, Russia). The
star was detected on 19 B band plates covering a 30-year period from
1965 to 1994. We found that over this period the B magnitude of
MN112 was $\simeq 17.2\pm 0.2$.

Although disappointing, the non-detection of large variability of
MN112 is not completely unexpected in view of similarity between
this star and the bona fide LBV P\,Cygni, whose brightness remains
relatively stable during the last three centuries and shows
irregular variability of $\sim 0.2$ mag superimposed on the gradual
brightening by $\simeq 0.1$ mag per century (de Groot et al. 2001).
Moreover, it is quite possible that a significant fraction of LBVs
(if not all of them) goes through the long quiescent periods
(lasting centuries or more; e.g. Lamers 1986) so that the fast
variability (on time scales from years to decades) observed in the
vast majority of classical LBVs could be merely due to the selection
effect (e.g. Massey et al. 2007).

\section{Conclusion and further work}

We identified a new Galactic cLBV via detection of a ring nebula
(reminiscent of the circumstellar nebula associated with the cLBV
GAL 079.29+00.46) and follow-up spectroscopy of its central star
showing that the spectrum of the star is almost identical to that of
the bona fide LBV P\,Cygni. Taken together, these finding strongly
suggest that the detected object, MN112, is a LBV. To unambiguously
prove the LBV nature of MN112, it is necessary to detect the major
changes in its brightness and spectrum. Although we realize that
MN112, like P\,Cygni, could be in the long-term quiescent phase, we
have launched a spectrophotometric monitoring of this star in the
hope that the "duck" will "quack" in the foreseeable future (cf.
Bohannan 1997). Confirmation of the LBV nature of MN112 will add
another object to the rare class of transitional massive stars and
will have profound consequences for understanding their evolution
and interaction with the ambient medium.

\section{Acknowledgements}

We thank the Calar Alto Observatory for allocation of director's
discretionary time to this programme and to the anonymous referee
for useful comments. VVG is grateful to his mother-in-law for
financial support. AYK acknowledges support from the National
Research Foundation of South Africa. SF and OS thank A.F.Valeev
for help with the observations and acknowledge support from the
RFBR grants N\,07-02-00909 and 09-02-00163. AMC acknowledges
support from the RFBR grant N\,08-02-01220 and the State Program
of Support for Leading Scientific Schools of the Russian
Federation (grant NSh-1685.2008.2). This work is based in part on
archival data obtained with the Spitzer Space Telescope, which is
operated by the Jet Propulsion Laboratory, California Institute of
Technology under a contract with NASA, and has made use of the
NASA/IPAC Infrared Science Archive, which is operated by the Jet
Propulsion Laboratory, California Institute of Technology, under
contract with the National Aeronautics and Space Administration,
the SIMBAD database and the VizieR catalogue access tool, both
operated at CDS, Strasbourg, France.

\end{document}